\documentclass[12pt]{article}
\usepackage{amsmath}
\usepackage{amssymb}
\usepackage{amsfonts}
\usepackage{amscd}

\newcommand{\R}{\mathbb{R}}

\newtheorem{theor}{Theorem}[section]

\def\bd{\begin{displaystyle}}
\def\ed{\end{displaystyle}}
\def\ba{\begin{eqnarray}}
\def\ea{\end{eqnarray}}
\def\bea{\begin{eqnarray*}}
\def\eea{\end{eqnarray*}}

\let\BBox\Box
\def\Box{$\BBox$}

\newcommand{\pr}{\textit{\textbf{Proof: }}}

\begin{document} 
\title{ Semi-groups and time operators for quantum  unstable systems \thanks{Presented at the
conference Group 25, session
"Semigroups, time asymmetry and resonances"
Cocoyoc, Mexico, 1-6 August, 2004.}}
\author{Maurice Courbage  }
\date{\today} 
\maketitle
\centerline{\it  Universit\'e Paris 7 - Denis Diderot }
\centerline{\it Laboratoire de Physique Th\'eorique  de la Mati\`ere Condens\'ee}
\centerline{\it  F\'ed\'eration Mati\`ere et Syst\`emes Complexes}
\centerline{\it Tour 24-14.5\`eme \'etage, 4, Place Jussieu}
\centerline{\it 75251 Paris Cedex 05 / FRANCE}\centerline{\it email:courbage@ccr.jussieu.fr}

\begin{abstract}
We use spectral projections of time operator in the Liouville space for simple quantum scattering systems in order to define a space of
unstable particle states evolving under a contractive semi-group. This space  includes  purely exponentially decaying
states that correspond to complex eigenvalues of this semi-group. The construction provides a  probabilistic interpretation
 of the resonant states characterized in terms of the Hardy class.

\end{abstract}

\section{Introduction}
\indent

We shall consider the unstable particle states of a quantum mechanical system in the framework  of the
Liouville von-Neumann time evolution. The decay processes are generally described by exponential time distributions 
with a given characteristic "lifetime".  As well known, such properties are unattainable within the standard quantum mechanics on
account of deviation from the exponential decay (see \cite{horwitz, misrasud, misrasudchiu} and
references therein to previous works ). Several attempts have been proposed to overcome these
difficulties and  it was proposed to obtain these pheneomena as resulting from semi-group evolutions
(\cite{ horwitz, bohm}). For example, It has been suggested by  Horwitz and Marchand 
\cite{horwitz} to describe the  decaying system by a subspace $
P \mathfrak{H}$ of $ \mathfrak{H}$ such that the family  $Pe^{-itH}P$ can be approximated by 
 a semi-group of contractive operators,  where  $P$ is some orthogonal projection. The semi-group is
associated to poles of the reduced resolvent $ P \frac{1}{H-z}P$ and the unstable space is generally defined according to
Weisskopf-Wigner
theory. That is,  briefly speaking,  the hamiltonian is represented as a sum of a free hamiltonian
$H_0$ and some interaction potential $gV$, where g is a coupling constant:  
\ba
H=H_0 + gV
\ea
and  the subspace $P\mathfrak{H}$ is taken as the space of eigenvectors of $H_0$ expected to decay under the total evolution group
$e^{-itH}$.  In that case, the decay law of the unstable particles,  which expresses the probability that the unstable states $\phi \in 
P \mathfrak{H}$,
created at time $t=0$, is still in the subspace $P\mathfrak{H}$ at time $t$, is given by:
\ba
p_{\phi}(t) =  \|
Pe^{-itH}\phi\|^2
\ea 

However,  the semi-group property does not hold and the family  $Pe^{-itH}P$  only obeys to a generalized
Master Equation deviating from the pure exponential decay on account of essential singularities of  the reduced resolvent
 (see e.g.
\cite{horwitz}, and  \cite{cour4} for a treatment of this aspect in the Liouville space).  

In this paper we study a semi-group
evolution obtained from time operators introduced in the Liouville space formulation of
quantum mechanics in \cite{mpc} and whose
existence and construction given in \cite{cour1}.

To start with a brief recapitulation, let  
$H$ be the Hamiltonian of the system acting on the Hilbert space $ \mathfrak{H}$. The "Liouville   
space", denoted $ \mathfrak{L}$, is the space of Hilbert-Schmidt  operators
$\rho $ on $
\mathfrak{H}$ such that $Tr(\rho^* \rho) <
\infty$,   equipped with the scalar product: $<\rho, \rho'> = Tr(\rho^* \rho ') $. The
time evolution of these operators is given by the  Liouville von-Neumann group of operators: 
\ba
U_t\rho = e^{-itH}\rho e^{itH} 
\ea 

The infinitesimal self-adjoint generator of this group is the Liouville von-Neumann operator L given by: 
\ba
L\rho = H\rho - \rho H
\ea

That is, $U_t = e^{-itlL}$. Some more mathematical details on this operator may be found, e.g. in  \cite{cour4}. 

 Here, instead of the evolution group
$e^{-itH}$ on 
$
\mathfrak{H}$ we shall consider the group
$ e^{-itL}$ on
$
\mathfrak{L}$ and we shall use as a projection operator P on the subspace of unstable states  the one associated to spectral
decomposition of time operator as will be explained in the section 2. 

It is well-known \cite{jamer} that  tentative constructions of a time operator for quantum machanical systems  leading to a rigourous 
understanding of the
fourth uncertainty relation faced the  remark of Pauli concerning the nonexistence of a 
canonically conjugated operator $T $ to the time evolution generator $H$, verifying:
$$
[H , T] = iI
$$
on account of the lower semidoundedness of the spectrum of the
hamiltonian. However, time operator was considered in the framework of the Liouville von-Neumann space and the fourth
uncertainty relation has been derived through it \cite{cour1}.  Here we shall restrict to this
framework. In section 3, we give a mathematical characterization of the subspace of unstable states 
which has the structure of the  Lax-Phillips theory \cite{lax}.

\section{Unstable States and time operators in the Hilbert-Schmidt space }

A
sufficient condition for the existence of a self adjoint time operator canonically conjugated to the Liouville operator, i.e.:
\ba
[L , T] = iI
\ea
 is that the hamiltonian has a lower 
semi-unbounded (absolutely) continuous spectrum. We shall suppose that $H$ has no singular
discrete or continuous spectrum. In the opposite case, $T$ should be defined on the orthogonal of
this singular subspace. The  construction  is recapitulated in the next section. Equation (2.5)  is
equivalent to  the Weyl relation: 
\ba
U_{-t} T U_{t} = T + tI
\ea

Denoting  by $ P_\tau$ the family  of spectral projection operators of T defined by: 
$$
T = \int_{\R} \tau dP_{\tau}
$$
 we have the following properties:\\

 i)  $P_\tau
  P_{\tau'} =   P_\tau$ if $\tau \leq \tau' $  ( a characteristic property of any spectral 
projection operators family) \\ 

 ii) $U_t P_\tau U_{-t} = P_{\tau + t}$ (an equivalent property of the Weyl relation).\\

It defines the family of
subspaces  ${\cal F}_\tau$, on which  projects $ P_\tau$, verifying:\\

 i)' ${\cal F}_{\tau} \subseteq {\cal F}_{\tau + t}$ for $ t \geq O
$ \\

 ii)' $U_t{\cal
F}_\tau = {\cal F}_{\tau + t}$. \\ 

 As proposed in \cite{cour2, cour3}, we associate the subspace ${\cal F}_{t_0}$ to the set of decaying
initial states  prepared at time
$t_0$. Shifting  the origin of time to the "time of preparation" $ t_0$ allows to consider, without loss
of generality, $P_0$ as a projection operator on the subspace of the unstable states. 

 In fact, two important properties are fullfilled by this subspace which enable it to realize such 
description.

 First,  it was proved that time operator satisfies to the fourth uncertainty relation between time and energy \cite{cour1} with the following
sense.  Considering  that a time operator should be  a quantum observable, like spatial position and
energy,  which describes the time occurrence of specified events such as time of arrival of a beam of
particles to a screen  or time of decay of unstable particles    and extending the von-Neumann
formulation of quantum mechanics to Liouville space, it is possible to define the states of a quantum
system by  normalized elements $\rho \in
\mathfrak{L}$ with respect to the scalar product, the expectation
of $T$ in the state
$\rho$ by: 
\ba
\langle  T \rangle _\rho = \langle \rho, T\rho\rangle 
\ea
and the "uncertainty" of the observable $T$ as its fluctuation in the state $\rho$:
\ba
(\Delta T)_\rho = \sqrt{\langle  T^2 \rangle _\rho - (\langle  T \rangle _\rho)^2}
\ea
 
Embedding  the normalized elements  $\psi \in \mathfrak{H}$ as 
elements $
\rho = 
\mid \psi \rangle \langle \psi \mid  \in \mathfrak{L}$, and the observables $A$ operating on $\mathfrak{H}$ 
as observables $\hat{A}$ operating on
$\mathfrak{L}$ as a multiplication by $A$: $\hat{A}\rho = A.\rho$, the above definition coincides with the usual quantum
rule giving the expectation of an observable $A$, operating on
$\mathfrak{H}$, in the state $\psi \in
\mathfrak{H}$ :
\ba
\langle  A \rangle _\rho = \langle \rho, \hat{A}\rho \rangle = \langle \psi, A\psi \rangle
\ea 
 A density matrix state  $M$ (i.e. a positive operator on 
$\mathfrak{H}$ with $Tr(M) = 1$), is embedded 
in
$\mathfrak{L}$ as  element $\rho = M^{1/2}$. Then, the expectation of the observable A  operating on $\mathfrak{H}$ in the mixture state  
$M$ usually given by $ Tr(M.A)$ is also preserved, for : 
\ba
\langle  A \rangle _\rho =  \langle \rho, \hat{A}\rho \rangle = Tr(M.A)
\ea
Let $\Delta E$ be the usual energy uncertainty in the state $M$ given by:
\ba
\Delta E =  \sqrt{Tr(MH^2) - (Tr(M.H))^2}
\ea
and  $ \Delta T = (\Delta T)_{M^{1/2}}$ be the uncertainty of $T$ in the state $M$ defined as in (2.8).  It has been shown
that:
\ba
\Delta E . \Delta T  \geq  \frac{1}{2 \sqrt{2}}
\ea
This  uncertainty relation leads to the interpretation of $T$ as the time occurrence of specified random events. The time of
occurrence of such events fluctuates and we speak of the probability of its occurrence  in a time interval $I = ]t_1, t_2] $. The
observable $T'$ associated to such event in the initial state $\rho_0$ has to be related to the time parameter $t $ by:
\ba
\langle  T' \rangle _{\rho_{t}} = \langle  T' \rangle _{\rho_{0}} -t
\ea
where $\rho_{t}= e^{-itL}\rho_{0}$. Comparing this condition with the above Weyl relation we see that
 we have to define  $T' $ as: $T'= -T$. Let $P_\tau
'$ be the family of spectral projections of $T'$, then, in the state $\rho$, the 
probability of ocurrence of the event in a time interval $I$ is given, as in the usual von Neumann formulation, by:
\ba
P(I,\rho)  =\| P'_{ t_2}\rho\|^2 - \| P'_{ t_1}\rho\|^2 = \| (P'_{ t_2} - P'_{ t_1})\rho\|^2 : =  \| P'( I)\rho\|^2
\ea

The  unstable "undecayed" states prepared at $t_0 = 0$ are the states  $\rho$ such that $P(I,\rho) = 0$ for any negative time interval $I$, that is:
\ba
\| P'_{ \tau}\rho\|^2 = 0, \forall \tau \leq 0
\ea 

In other words, these are the states verifying $P'_{ 0}\rho = 0$. It is straightforwordly checked that   the spectral projections  $P'_{ \tau}$
are related to the spectral projections
$  P_{
\tau}$ by the following relation:
\ba
P'_{ \tau}= 1- P_{- \tau}
\ea
Thus, the unstable states are those states verifying:
$\rho = P_{ 0}\rho$ and they coincide with our subspace ${\cal F}_{0}$. For these states, the probability that a system
prepared in the undecayed state
$\rho$ is found to decay sometime during the interval $I = ]0, t]$ is $\| P'_{ t}\rho\|^2 = 1 - \| P_{ - t}\rho\|^2$, a monotonically nondecreasing quantity
which converges to 1 as
$t
\rightarrow
\infty$ for $\| P_{ - t}\rho\|^2$ tends monotonically to zero.  As noticed by Misra and Sudarshan \cite{misrasud}, such
quantity could not exist in the usual quantum mechanical treatment of the decay processes and could not be related to  the
"survival probability" (1.2) for it is not a monotonically decreasing quantity in the Hilbert space formulation. In the Liouville
space, given any intial state
$\rho$, its survival probability in the unstable space is given by:
\ba
p_{\rho}(t) = \| P_0e^{-itL}\rho \|^2
\ea  

This survival probability and the probability of finding  the system to decay sometime during the interval $I = ]0, t]$, are
related by:
\ba \nonumber
 \| P'_{ t}\rho\|^2&= &1 - \| P_{ - t}\rho\|^2 \\ \nonumber
&= &1 -  \| U_{-t}P_0 U_t\rho \|^2\\ \nonumber
& = &1 - \| P_0e^{-itL}\rho \|^2 \\
& =& 1 - p_{\rho}(t)
\ea 
The survival probability  is  monotonically decreasing to 0 as $t \rightarrow \infty$.  This is true for 
for any general intial state as ca be seen from the equation (2.18). It should noted that the projection
operator $ P_0$ is not a "factorizable"    operator, that is , not of the form $P_0\rho = E \rho E$ where
E is a projection operator. 

Second, the projection  
$P_{t_0}\rho_t$ 
 obeys to a closed  equation (i.e. 
it depends only on the projected  initial condition
$P_{t_0}\rho(0)$) given by a contraction semi-group for $t>t_0$. This is a  consequence of the properties i) and ii) leading for any $t>0$ to:
$$
P_{t_0} = P_{t_0}P_{t_0+t} \\
= P_{t_0}U_{t }P_{t_0}U_{-t }
$$
Thus, multiplying both members by  $U_{t} $, we obtain: 
\ba
P_{t_0}U_{t} = P_{t_0}U_{t }P_{t_0} 
\ea
 It 
implies that the one parameter family of operators: 
\ba
W_t = P_{t_0}U_{t} = P_{t_0}U_{t }P_{t_0}
\ea
is a semi-group for $t>0$:
\ba
W_t W_{t'} =  W_{t+ t'}
\ea
for any $t, t' >0$. It is also evident that this is a contractive semi-group, i.e., $\| W_t\|  \leq 1$, with respect to the operator norm on $
\mathfrak{L}$.  Note that equation (2.10) is nothing but the fact that the  space  ${\cal F}_{t_0}^\bot$, orthogonal to ${\cal F}_{t_0}$, is
invariant under
$U_{-t }$.  
 
 The above semi-group has quite  analog structure than the Lax-Phillips semi-group \cite{lax}. For more informations on the
applications of this structure,  we refer to the paper of Y. Strauss \cite{strauss} and references therein . \\

\section{Mathematical characterization of $P_0$ and $W_t$}

  In this section, we shall restrict to the case where the hamiltonian $H$ has a simple Lebesgue spectrum extending from $0$ to
$+\infty$. Some models, like Friedrichs models \cite{fried}, verify this assumption  which is useful in order to illustrate the description of
unstable states in the Hilbert-Schmidt space. 

Choosing a spectral representation of $H$, any $\psi \in   \mathfrak{H}$ is represented by a
 square integrable
function $\psi(\lambda) \in L^2(\R^+)$ and  the hamiltonian is represented by the multiplication operator by $\lambda$, that is, $H\psi(\lambda) =
\lambda\psi(\lambda)$. The Hilbert-Schmidt operators on $ L^2(\R^+)$ are the integral operators associated to
square-integrable kernels, $\rho(\lambda, \lambda') \in  L^2(\R^+ \times \R^+)$. The Liouville operator  is given by: 
$L \rho(\lambda, \lambda') = (\lambda- \lambda')  \rho(\lambda, \lambda')$. Now the spectral representation of L is obtained through the
change of variables: $(\lambda, \lambda') \rightarrow (\nu, E) $ defined by:
\ba
\nu &= & \lambda- \lambda' \\
E &= & max(\lambda, \lambda') 
\ea 

In  the spectral representation of $L$, we denote again, for the sake of simplicity, a Hilbert-Schmidt operator on
 $\mathfrak{H}$ by $\rho(\nu, E) \in 
L^2(\R \times \R^+)$, thus :
\ba
L \rho(\nu, E) = \nu \rho(\nu, E) 
\ea
Time operator $T\rho(\nu, E)$ is then  the self-adjoint extension of  the operator $i
\frac{\partial}{\partial \nu}\rho(\nu, E)$. Under this hypothesis on the Hamilton operator we obtain a spectral representation of
$T$ using the Fourier Transform $$\hat{\rho}(\tau, E) = \frac{1}{\sqrt{2 \pi}} \int_{-\infty}^{+\infty} e^{i\tau \nu}\rho(\nu,
E)d\nu$$
In this representation, the time operator  will be given by: $T\hat{\rho}(\tau, E) = \tau
\hat{\rho}(\tau, E)$ and the spectral projection operator $P_\tau$ is the multiplication operator by the
characteristic function of $]-\infty, \tau ]$ denoted $ \chi_{]-\infty,
\tau ]}$. It follows from the Paley-Wiener theorem that the subspace of unstable states ${\cal F}_{0}$ is characterized as the boudary values on
$\R$ of the upper Hardy class of complex functions 
$\mathfrak{H}^+$ defined as the  functions $\rho(z,E)$ that are upper-plane analytic vector valued  of $z \in \mathbb{C}_+ =
\{z
\in 
\mathbb{C}: Im(z)>0\}$ such that: 

\ba
sup_{y>0} \int_{-\infty}^{+\infty}\|\rho(x+iy,E)\|_E dy < \infty
\ea
where $ \|\rho(x+iy,E)\|^2_E = \int_{0}^{+\infty}|\rho(x+iy,E)|^2 dE $. It is also clear that the complementary  orthogonal  projection
operator is the multiplication operator by the charcteristic function of $]0, +\infty [$ on the space of the boudary values on
$\R$ of the lower Hardy class
$\mathfrak{H}^-$.  It follows that a state  $\rho(\nu, E) \in  \times \R^+)$ belongs to ${\cal F}_{0}$ if and only if $\rho(\nu, E)$
belongs to
 the Hardy space $  \mathfrak{H}^+$. As $ \mathfrak{L} = \mathfrak{H}^+  \oplus \mathfrak{H}^-$, any initial state $ \rho$ is decomposed into a sum
 $ \rho = \rho^+ +
\rho^-$ of unstable  component $\rho^+ \in \mathfrak{H}^+$ and some orthogonal component $
\rho^- \in \mathfrak{H}^-$. 

Let us note that we used a slight generalization of the Paley-Wiener theorem. The original theorem was fromulated in $L^2(\R)$.
But we can reduce the space $L^2(\R \times \R^+)$ to a countable direct sum of copies of $L^2(\R)$, by taking a basis $\phi_k
(E)$ of $L^2( \R^+)$ and expanding $\rho(\nu, E)$ on this basis:  $\rho(\nu, E) = \sum_{k = 0}^\infty a_k(\nu) \phi_k
(E), a_k(\nu) \in L^2(\R)$. Any $\rho(\nu, E) \in L^2(\R \times \R^+)$ is then identified with the
 corresponding sequence of functions $ \lbrace a_k(\nu) \rbrace $, for each of its elements the Paley-Wiener theorem applies. The
same argument should be understood in what follows. \\

For a general state the decay rate of the survival probability is not exponential. However, there is a family of
unstable intial states whose evolution under the semi-group $W_t$ has pure exponential decay: 

\begin{theor}
For any $\xi \in
\mathbb{C}_- =
\{z
\in 
\mathbb{C}: Im(z)<0\}, e^{-it\xi}$ is an eigenvalue of the semi-group $W_t$  corresponding to the eigenfunctions: 
\ba
\rho_\xi(\nu, E) = \frac{\psi (E)}{\nu - \xi}, \, \,  \psi \in  L^2( \R^+)
\ea. 
 
\end{theor}

\noindent \pr
The function $\rho_0(\nu, E) = \frac{\psi (E)}{\nu - \xi}, \in \mathfrak{H}^+$, so that $\rho_0 = P_0 \rho_0$. Thus, in
the spectral representation of $L$, $\rho_t(\nu,E) := U_t \rho_0(\nu,E) = \frac{e^{-it\nu }\psi (E)}{\nu -
\xi}$, and denoting by $\hat{\rho_t}(\tau, E) = \frac{1}{\sqrt{2\pi}} \int_{-\infty}^{+\infty}e^{i\tau
\nu}\rho_t(\nu,E)$ the Fourier transform of $\rho_t$ with respect to $\nu$, we obtain for $\tau
<0$ and $t>0$, $ P_0\hat{\rho_t}(\tau, E) = 
\hat{\rho_t}(\tau, E) = -\sqrt{2\pi}i e^{i(\tau - t)\xi}\psi(E)$,   and $ P_0\hat{\rho_t}(\tau, E)=0 $ for $\tau >0$. Taking
the inverse Fourier transform of $ P_0\hat{\rho_t}(\tau, E)$ we obtain in the spectral representation of the Liouville operator:

\ba 
W_t\rho_0(\nu,E) = P_0U_t \rho_0(\nu,E)  =  \frac{e^{-it\xi }\psi (E)}{\nu - \xi}
\ea
for $t>0$. $\blacksquare$  

It follows that these states have purely exponentially decaying survival probability. It is also a result of the equation (2.20) and the
above theorem that initial states with unstable part of the form (3.26) will also have purely exponentially decaying survival
probability. 
 
\section{Some concluding remarks }
\indent

The semi-group  derived from time operator is analog to the semi-group derived from 
Lax-Phillips evolution. Thus, Liouville space provides a natural realization of the Lax-Phillips  structure and a definition of the
resonances states. The above  eigenfunctions of
$W_t$ are the analog of the Gamov states much studied recently in the frame of rigged Hilbert space \cite{bohm}. But, the
introduction of a time operator allows to give them a probabilistic content. 

  We shall apply the above construction in a forthcoming
paper to some simple scattering models extended in the Liouville space, like the Friedrichs model. Recently, the
expectation of time operator in  the free hamiltonian eigenfunctions state of this model has been  computed 
and it has been shown that it coincides with the lifetime of  the resonance \cite{ordonnez}.

\end{document}